\documentclass{Interspeech2024}

\interspeechcameraready

\title{Speech-based Clinical Depression Screening: An Empirical Study}

\name[affiliation={1*}]{Yangbin}{Chen}
\name[affiliation={2*}]{Chenyang}{Xu}
\name[affiliation={1}]{Chunfeng}{Liang}
\name[affiliation={3}]{Yanbao}{Tao}
\name[affiliation={2\dagger}]{Chuan}{Shi}

\address{
  $^1$Suzhou Fubian Medical Technology Co., Ltd, China\\
  $^2$Peking University Sixth Hospital, China \\
  $^3$The First Clinical College of Xinxiang Medical University, China}
  
\email{dongyiwu92@gmail.com, 19801181160@163.com, cfliang666@gmail.com, 707839227@qq.com, shichuan@bjmu.edu.cn}


\begin{document}

\maketitle

\def\thefootnote{}\footnotetext{$\dagger$ denotes corresponding author; * denotes equal contribution.}
\begin{abstract}
    
    This study investigates the utility of speech signals for AI-based depression screening across varied interaction scenarios, including psychiatric interviews, chatbot conversations, and text readings.
    Participants include depressed patients recruited from the outpatient clinics of Peking University Sixth Hospital and control group members from the community, all diagnosed by psychiatrists following standardized diagnostic protocols.
    We extracted acoustic and deep speech features from each participant's segmented recordings.
    Classifications were made using neural networks or SVMs, with aggregated clip outcomes determining final assessments.
    Our analysis across interaction scenarios, speech processing techniques, and feature types confirms speech as a crucial marker for depression screening.
    Specifically, human-computer interaction matches clinical interview efficacy, surpassing reading tasks.
    Segment duration and quantity significantly affect model performance, with deep speech features substantially outperforming traditional acoustic features.
    
\end{abstract}

\section{Introduction}

\begin{figure*}[t]
  \centering
  \includegraphics[width=\linewidth]{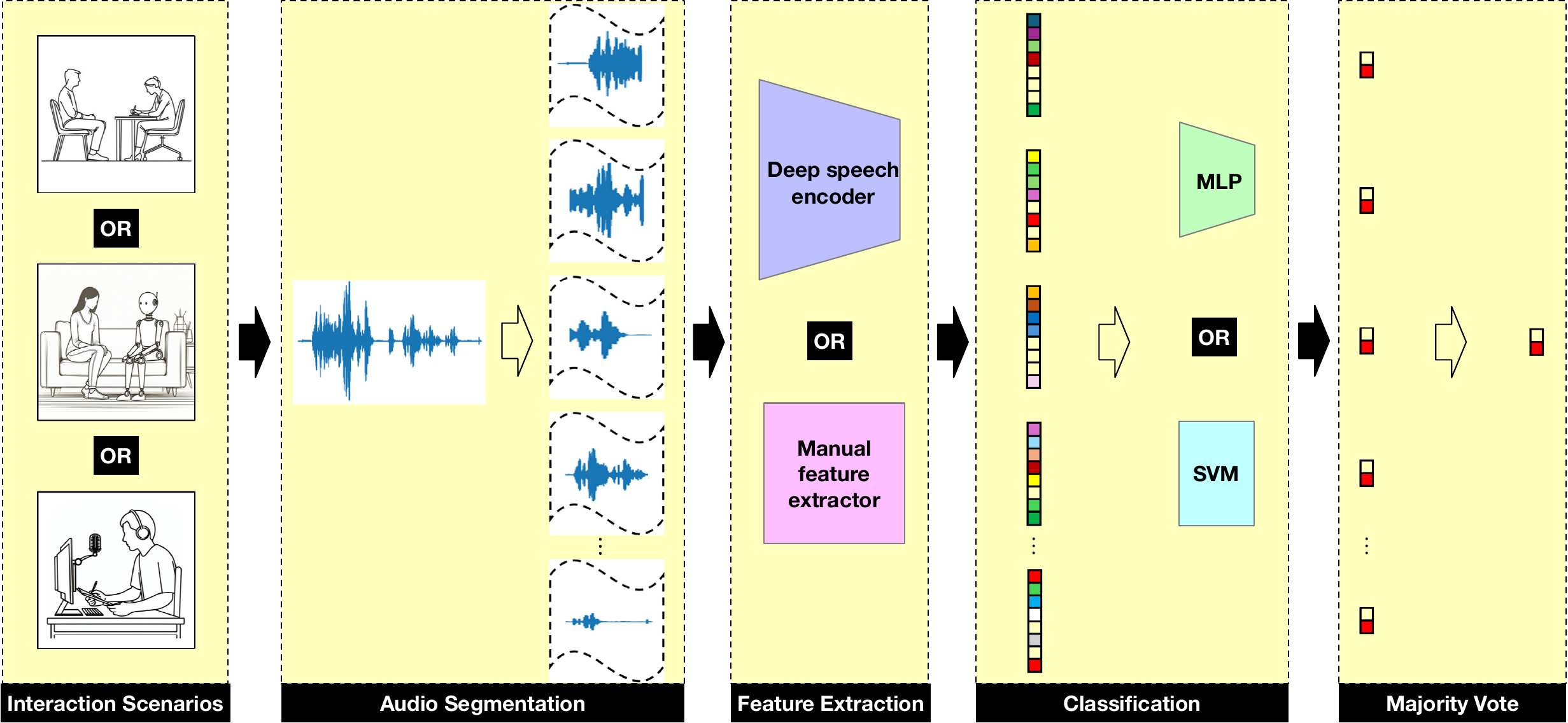}
  \caption{Framework of this study which consists of several stages: (1) Interaction scenarios -- psychiatric interviews, chatbot conversations, and text readings; (2) Audio segmentation -- to segment participants' recordings into audio clips; (3) Feature extraction -- to extract acoustic or deep speech features; (4) Classifier -- to do classification with simple MLP or SVM models; (5) Majority vote -- to determine the final prediction for each participant through voting among their clip outcomes.}
  \label{fig:framework}
\end{figure*}

Depression, recognized as a pervasive mental health disorder, afflicts around 300 million individuals globally \cite{nagy2020single}. 
Specifically, in China, adult lifetime depression prevalence stands at 6.8\% \cite{huang2019prevalence}.
Despite its prevalence, numerous barriers hinder effective depression management, including limited awareness, disparate access to healthcare, and variable service quality. 
Such barriers often lead to delayed screenings and diagnoses, underscoring the critical need for developing efficient, accessible, and affordable depression assessment tools and methods.

Recent advances in depression screening have increasingly leveraged artificial intelligence (AI), exploring diverse areas including electronic health record (EHR) mining \cite{sudhanthar2015improving}, biomarker identification \cite{ringeval2019avec,de2019depression,flint2021systematic}, daily activities monitoring \cite{smagula2022association}, and social media analysis \cite{skaik2020using}.
Among these, biomarker analysis offers an objective method.
Multiple hypotheses on depression exist, yet none fully explain its pathology, leaving a gap in clinical diagnostic markers \cite{cui2024major}.
Behavioral biomarkers, particularly speech signal analysis, are increasingly researched due to its advantages: it is non-invasive, user-friendly, and highly portable.

Research on speech-based depression screening encompasses several key domains: statistical analysis to explore correlations between depression and acoustic features like shimmer, F0, and MFCC \cite{wang2019acoustic}; adopting speech features to develop advanced deep neural networks for depression severity assessment \cite{seneviratne21_interspeech}; combining speech, visual and other data for multimodal depression detection \cite{fara23_interspeech}; and developing screening techniques for personalized depression diagnosis \cite{campbell23_interspeech}.

However, many studies face challenges, limited by the quality and annotation of samples. 
Unstandardized data collection processes in non-clinical settings compromise data integrity.
Relying on self-assessment scales to label depressed patients and overlooking the clinical relevance of reported symptoms lead to high false positive rates and inaccuracies.
Moreover, the scarcity of data coupled with the complexity of models exacerbates overfitting, undermining model generalizability.

This work explores and implements various strategies to enhance the effectiveness of AI-based depression screening models using speech signals (see Figure \ref{fig:framework}).
We collaborated with Peking University Sixth Hospital to recruit participants.
Participants currently experiencing a major depressive episode were clinically diagnosed with depression by psychiatrists in an outpatient setting, and confirmed to have no other past or current psychiatric disorders. 
To establish a control group, community volunteers who were not taking psychiatric medications and reported no symptoms of psychiatric disorders or other clinical conditions in the previous period were recruited. 
The Mini International Neuropsychiatric Interview (MINI) \cite{sheehan1998mini}, a tool commonly utilized for diagnosing psychiatric conditions, was administered to verify their diagnostic status.

We designed three interaction scenarios to collect speech data: psychiatric interviews \cite{xu2019automated}, chatbot conversations \cite{weisenburger2024conversational}, and text readings \cite{zhao2022vocal}.
Recordings from each participant were segmented into fixed-length clips, with adjustments to these clips' duration and quantity for comprehensive experimental analysis.
Ultimately, we assessed the impact of three traditional acoustic feature sets and a deep speech feature, utilizing simple neural networks or SVMs for classification.

Our main contributions can be summarized as follows:
\begin{itemize}
    \item Clinical data were collected from Peking University Sixth Hospital, incorporating rigorous inclusion criteria.
    Diverse data sets were obtained through various interaction scenarios.
    \item The study reveals that depression screening models trained on speech data from chatbot conversations are comparable to those trained on data from psychiatric interviews, both outperforming models trained on reading task data.
    \item We observed that with the increase in duration and number of audio clips per participant in training and testing, model performance enhances, albeit not in a linearly consistent manner.
    \item Our research demonstrates that deep speech features significantly surpass traditional acoustic features in depression screening, marking a substantial advancement in the field.
    \item In experiments with speech data from 270 participants, we obtained optimal results exceeding 90\% across metrics such as accuracy and specificity.
\end{itemize}

\section{Data collection}

\subsection{Standard procedures}
This study was ethically approved and required informed consent from all participants, whose data were strictly confidential.
We recruited individuals from a medical center.
Under the guidance of psychiatrists and technicians, participant completed assessments including medical histories, psychoactive substance use surveys, PHQ-9 self-evaluations, the MINI, the chatbot interviews, and text reading tasks.
The psychiatrists provided a depression diagnosis for each participant, categorizing them as either depressed or healthy.

\subsection{Interaction scenarios}



Participants' speech data were gathered in three distinct scenarios: psychiatric interviews, chatbot conversations, and text readings.
Psychiatric interviews are from the MINI by trained psychiatrists.
Chatbot conversations are from the AI diagnostic module, which is built on a custom dialogue system.
It employs large language models (LLMs) and the Hamilton Rating Scale for Depression (HAM-D) to query participants about depresive symptoms, offering empathetic responses, and topic-focused inquires based on their answers.
For the text readings, participants were asked to read a neutral passage aloud, which was recorded.
All participants read the same text, with an average task duration of about one minute.

\subsection{Data characteristics}
We build an experimental dataset from 270 participants.
Table \ref{tab:demo_charact} outlines the demographic characteristics of the datasets, showing no significant differences in age, gender, and educational years between depressed and healthy groups.
The dataset's strength is its demographic alignment. 
Since age and gender significantly impact vocal characteristics, their effects are controlled to minimize experimental bias. 
Notably, within identical interaction scenarios, the depressed group exhibited longer speech duration compared to their healthy counterparts.
Conversely, during reading tasks, the depressed group's average duration was shorter, attributed to their tendency to demonstrate fatigue and a desire to discontinue the task, prompting early termination of recordings.

\begin{flushleft}
\begin{table}[h]
    \centering
    \begin{tabular}{@{}lcc@{}}
    \toprule
         & Depressed(n=152) & Healthy(n=118) \\
    \midrule
        Age(years) & 31.7 $\pm$ 11.4 & 31.0 $\pm$ 9.9\\
        Gender(n) &  & \\
         \hspace{3mm}Female &  101 & 82 \\
         \hspace{3mm}Male &  51 & 36 \\
        Education years (n) &  & \\
         \hspace{3mm}$\leq$12 & 50 & 27 \\
         \hspace{3mm}$>$12 & 102 & 91 \\
        Audio duration (seconds) & & \\
         \hspace{3mm} Psychiatric interview & 242.6 & 72.1 \\
         \hspace{3mm} Chatbot conversation & 302.8 & 103.4 \\
         \hspace{3mm} Reading task & 53.1 & 60.0 \\
    \bottomrule
    \end{tabular}
    \caption{Demographic characteristics of all participants.}
    \label{tab:demo_charact}
\end{table}
\end{flushleft}

\begin{table*}[h]
    \centering
    \begin{tabular}{lccccccc}
    \toprule
         \textbf{Interactive mode} & \textbf{Duration} & \textbf{\#Audio clips } & \textbf{Feature} & \textbf{Accuracy} & \textbf{Sensitivity(Recall)} & \textbf{Specificity} & \textbf{Precision} \\
    \midrule
         Psychiatric interview & 5s & 5 & chinese-hubert & 90.9\% & 94.1\% & \textbf{86.7\%} & \underline{90.4\%}\\
         & 5s & 11 & chinese-hubert & \textbf{91.9\%} & 91.6\% & \underline{85.8\%} & \textbf{90.8\%}\\
         & 10s & 5 & chinese-hubert & \underline{91.5\%} & \underline{96.8\%} & 84.8\% & 89.2\%\\
         & 10s & 5 & eGeMAPSv02 & 77.6\% & 91.6\% & 59.5\% & 74.4\%\\
         & 10s & 5 & ComParE\_2016 & 86.0\% & 94.8\% & 74.7\% & 83.2\%\\
         & 10s & 5 & IS09-13 & 84.2\% & \textbf{97.4\%} & 67.2\% & 80.0\%\\
    \midrule
         Chatbot conversation & 5s & 5 & chinese-hubert & 93.3\% & \underline{96.7\%} & 88.8\% & 91.9\%\\
         & 5s & 11 & chinese-hubert & \underline{93.4\%} & \textbf{96.8\%} & \underline{89.0\%} & \underline{92.1\%}\\
         & 10s & 5 & chinese-hubert & \textbf{94.1\%} & \underline{96.7\%} & \textbf{90.7\%} & \textbf{93.2\%}\\
         & 10s & 5 & eGeMAPSv02 & 79.2\% & 91.5\% & 63.3\% & 76.7\%\\
         & 10s & 5 & ComParE\_2016 & 86.6\% & 92.1\% & 79.6\% & 85.8\%\\
         & 10s & 5 & IS09-13 & 86.2\% & 94.7\% & 75.3\% & 83.2\%\\
    \midrule
         Reading task & 5s & 5 & chinese-hubert & \underline{81.3\%} & \textbf{85.4\%} & 76.3\% & 82.9\%\\
         & 5s & 11 & chinese-hubert & 80.3\% & 82.2\% & \underline{77.9\%} & \underline{83.2\%}\\
         & 10s & 5 & chinese-hubert & \textbf{84.1\%} & \underline{84.8\%} & \textbf{83.0\%} & \textbf{86.6\%}\\
         & 10s & 5 & eGeMAPSv02 & 72.4\% & 77.5\% & 65.7\% & 74.6\%\\
         & 10s & 5 & ComParE\_2016 & 75.9\% & 80.2\% & 70.2\% & 77.4\%\\
         & 10s & 5 & IS09-13 & 69.6\% & 77.5\% & 59.2\% & 71.1\%\\
    \bottomrule
    \end{tabular}
    \caption{Overall results of speech-based machine learning models across different interaction scenarios, speech segment processing techniques, and speech feature types. The bold and underlined results indicate the best and second-best performances, respectively.}
    \label{tab:results}
\end{table*}

\subsection{Experimental settings}
Utilizing the collected speech data, we trained several machine learning models and conducted comprehensive and detailed experimental analyses across different interaction scenarios, speech segment processing techniques, and speech feature types.
Each experiment was conducted using 5-fold cross-validation, partitioning the data by individual participants to guarantee that data from the same participant were not simultaneously included in both the training and testing sets.
All experiments were executed on a computer equipped with a NVIDIA GeForce RTX 3060 card.
The overall results are presented in Table \ref{tab:results}.

\section{Interaction scenario analysis}
Depression screening based on speech data has been extensively studied, using various data collection methods, including: (1) collecting data from offline face-to-face interviews or online remote consultations between doctors and patients; (2) gathering data through interactive tasks like text reading, picture description, and video-based question answering; (3) acquiring data via interactions with chatbots.
Different interaction modes may influence participants' mindset and behaviours, especially when the collected data is intended for biomarker detection.
Few studies in AI-based depression screening have examined the impact of varying interaction methods on individuals.
Notably, our study identified that some depressed participants might exhibit impatience during reading tasks in later stages.
Furthermore, data collection processes that provoke pessimistic emotions in depressed patients through negative topics are considered inappropriate from a humanitarian perspective.
Therefore, this work specifically analyzes the concept of interaction, aiming to provide new insights and evidence for developing a user-friendly depression screening method.

In this work, within the collected data across three scenarios, speech from psychiatric interviews, containing full dialogues, required participants' voices to be separately extracted.
In contrast, their voices from chatbot conversations and reading tasks were automatically saved without other voices.
For each participant, we employed a sliding window approach with a $T$-second duration to randomly select $N$ speech segments from their entire recordings.

\begin{figure}[h]
  \centering
  \includegraphics[width=\linewidth]{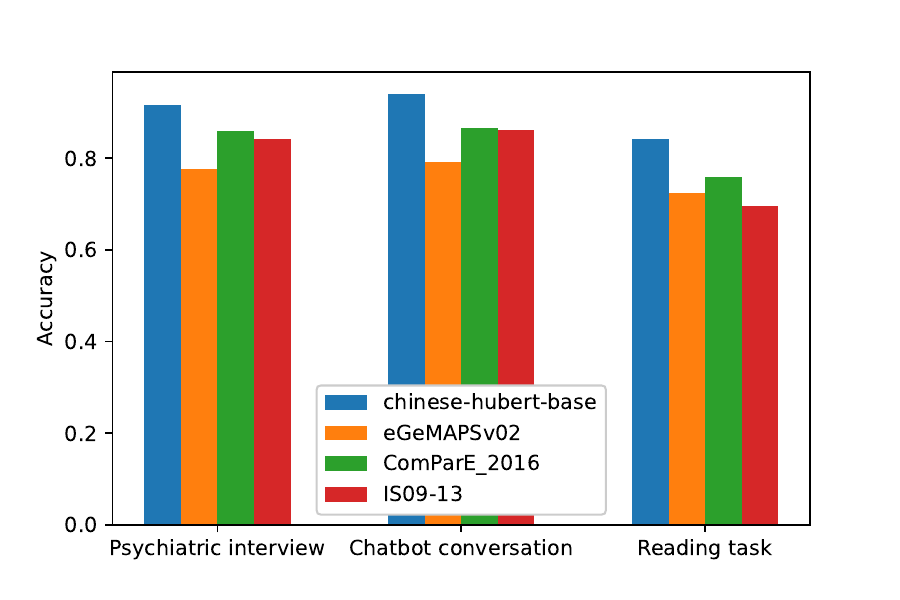}
  \caption{Performance comparison of using different audio features across three interaction scenarios.}
  \label{fig:Scenario_comparison}
\end{figure}

Figure \ref{fig:Scenario_comparison} displays the results of training classifiers with various interaction scenarios, configured with $T$ set to 10 and $N$ set to 5.
Results from Table \ref{tab:results} and Figure \ref{fig:Scenario_comparison} indicate that speech data from human-computer interaction scenarios, when utilized for model training, achieved optimal test performance, surpassing 90\% in accuracy, sensitivity, specificity, and precision.
Similarly, models trained with speech data from psychiatric interviews exceeds 90\% across these metrics, except for specificity.
However, models trained on reading task speech data demonstrated optimal metrics ranging from 80\% and 90\%.

\textcolor{blue}{\textbf{SUMMARY}: AI-based depression screening models trained with speech data demonstrate overall excellent performance.
Notably, models utilizing speech data from human-computer conversations and face-to-face psychiatric interviews reach comparable or superior outcomes.
In contrast, models based on speech data from text reading tasks showed lower efficacy than those trained with the former two scenarios.}

\section{Speech processing analysis}
Our methodology for processing speech data posits that the vocal manifestations of individuals with depression tend to maintain consistency across contexts. 
Consequently, this study adopts a uniform approach to evaluating an individual's vocal performance throughout the experimental procedure. 
As outlined in Section 3, we employ a methodology that involves using a $T$-second window to extract $N$ speech segments for use as either training or testing data. 
Ultimately, the diagnosis of depression is inferred by aggregating the model's predictions across all speech segments of a participant through a voting mechanism in the testing phase.

\begin{figure}[h]
  \centering
  \includegraphics[width=\linewidth]{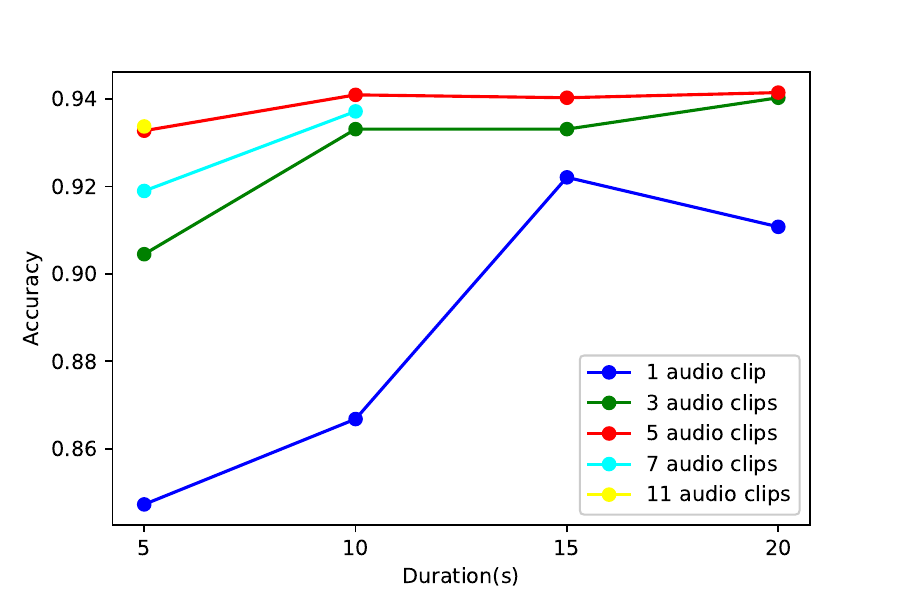}
  \caption{Performance comparison of using different numbers and duration of audio clips from individuals.}
  \label{fig:clips_analysis}
\end{figure}

We investigated the impact of $T$ and $N$ values on model performance using speech data from the chatbot conversation scenario, with deep speech features as model inputs. 
$T$ values were set at 5, 10, 15, and 20 seconds, while $N$ values were chosen as 1, 3, and 5. 
Given the constraints of total audio duration, extracting 7 pieces of 15-second or 20-second segments from each participant's audio presented challenges.
Hence, with $N$ set to 7, $T$ was capped at 10 seconds. 
Additionally, we explored a unique set of values, $T=5$ and $N=11$, to assess the effects of larger $N$ values on outcomes. 
The experimental results are illustrated in Table \ref{tab:results} and Figure \ref{fig:clips_analysis}.

Upon examining the impact of speech segment length (T), model efficacy improves when T is extended from 5 to 10 seconds in all experimental conditions. 
Further extension of T to 15 seconds significantly enhances model performance for a single segment sampling, whereas the effect is less pronounced for sampling three and five segments. 
An increase of T to 20 seconds results in diminished outcomes for a single segment, improvement for three segments, and negligible change for five segments.
In the analysis of speech segment count (N), an overall enhancement in model outcomes is observed with the increment of N, regardless of segment duration. 
However, a decline in model performance is noted when N increases to 7 with a 5-second duration, suggesting that such a brief duration may not adequately capture consistent vocal biomarkers indicative of depression, potentially due to the influence of anomalous situations. 
When N is adjusted to 11 for a 5-second duration, the performance remains comparable to that with N set at 5.

\textcolor{blue}{\textbf{SUMMARY}: Sampling speech segments acts as a data augmentation strategy, where the duration and quantity of segments considerably influence the performance of the model. 
Enhancements in these parameters generally lead to improved model outcomes.
However, the correlation between these increases and model performance is not linear.}

\section{Feature type analysis}
Feature extraction plays an essential role in AI tasks, particularly within the domain of medical AI applications. Speech-based depression screening frequently adopts conventional feature sets designed for speech emotion recognition.
The eGeMAPSv02, an extended version of the Geneva Minimalistic Acoustic Parameter Set, is a refined acoustic feature set tailored for affective computing \cite{eyben2015geneva}. 
It encompasses a broad spectrum of features capturing frequency, energy, spectral, and temporal characteristics of speech. 
The ComParE\_2016 feature set, developed for the Computational Paralinguistics Challenge, includes 6,373 attributes covering spectral, prosodic, and voice quality aspects\cite{schuller2016interspeech}.
Another feature set experimented in our work is IS09-13, which originates from the Interspeech 2009 Emotion Challenge, providing a comprehensive framework for analyzing emotional expressions in speech\cite{schuller2009interspeech}.
All these feature sets have been widely utilized in emotion recognition, stress detection, and health-related speech analysis.

HuBERT (Hidden Unit BERT) is a self-supervised learning model for speech representation, which enhances speech processing by predicting hidden units of speech segments, which are clustered from raw audio without relying on labeled data.
Pre-training involves masked prediction of these units, enabling HuBERT to capture rich acoustic and linguistic features. \cite{hsu2021hubert}. 
We leverage its Chinese version \footnote{\url{https://github.com/TencentGameMate/chinese_speech_pretrain}} to compare its performance with above three conventional feature sets \cite{eyben2010opensmile}.
From Table \ref{tab:results} and Figure \ref{fig:Scenario_comparison}, we find that the chinese-hubert features outperforms other three feature sets in all scenarios.
The traditional acoustic features perform good in terms of sensitivity but underperform in other metrics.

\textcolor{blue}{\textbf{SUMMARY}: Deep speech features extracted from large pre-trained models are significantly useful in downstream tasks like depression screening, even with a simple classifier.}

\section{Conclusion}
Our study analyzes speech-based depression screening from three dimensions: interaction scenarios, speech processing techniques, and feature types, with the goal of identifying optimal practices. 
We find that human-computer interaction platforms are as effective as direct psychiatric interviews, highlighting the potential for simplifying and standardizing depression diagnosis and monitoring, which could enhance accessibility and consistency in assessment.
Selecting appropriate duration and numbers of speech segments can improve model performance. 
Moreover, Deep speech features surpass traditional acoustic features even when utilizing basic classifiers.
In the future, we can focus on developing an efficient, accessible, and user-friendly tool for depression screening, leveraging these insights.
Furthermore, we will dedicate additional efforts on automated depression diagnosis methods that can more precisely conduct depression severity assessment.

\section{Statement of using AI-assisted tools}

During the preparation of this work, the authors used ChatGPT-4 only for language polishing in order to enhance the clarity and readability of the text. After using this tool, the authors reviewed and edited the content as needed and took full responsibility for the content of the publication.

\bibliographystyle{IEEEtran}
\bibliography{template}

\end{document}